\documentclass[a4paper]{article}

\usepackage[dvips]{graphicx}
\usepackage{bm}
\usepackage{cite}

\def\d{{\rm d}}

\makeatletter

\def\equation{$$ 
  \stepcounter{equation}\let\@currentlabel=\@eqnnum}

\def\eqnarray{\stepcounter{equation}\let\@currentlabel=\@eqnnum
\global\@eqnswtrue
\global\@eqcnt\z@\tabskip\@centering\let\\=\@eqncr
$$\halign to \displaywidth\bgroup\@eqnsel\hskip\@centering
  $\displaystyle\tabskip\z@{##}$&\global\@eqcnt\@ne 
  \hfil$\;{##}\;$\hfil                                     
  &\global\@eqcnt\tw@ $\displaystyle\tabskip\z@{##}$\hfil  
   \tabskip\@centering&\llap{##}\tabskip\z@\cr}

\makeatother

\begin{document}
\title{
  Phase-locked and phase drift solutions 
  of phase oscillators
  with asymmetric coupling strengths
}
\author{
  YAMADA Hiroyasu
  \thanks{A visiting researcher at 
    R-laboratory,
    Department of Physics, Nagoya University, 
    Nagoya 464--8602, Japan}
  \thanks{A visiting researcher at 
    Bio-mimetic Control Research Centre, 
    The Institute of Physical and Chemical Research (RIKEN), 
    Nagoya 463--0003, Japan}
}
\date{
}

\maketitle

\begin{abstract}
  Phase-locked solutions of coupled oscillators are studied
  with asymmetric coupling strengths or inhomogeneous natural
  frequencies.
  The solutions show remarkable profiles of phase lags from the
  pacemaker corresponding to the ratio of upward and downward coupling
  strengths.
  By means of the existence condition of phase-locked solutions,
  the transition points from phase-locked to phase drift states are
  estimated.
  The application of the existence condition to the case of the
  linear gradient of natural frequency illustrates some scaling
  properties in the frequency diagrams.
\end{abstract}

\section{Introduction}
\label{sec: intro}

Coupled oscillators have been dealt with in wide and diverse areas
associated with oscillation and synchronization phenomena.
The phenomenon of collective synchronization of phases or frequencies
attracts attentions in not only biology 
but also in physics and engineering\cite{Winfree80,Kuramoto84}.
Theoretical studies have been deeply carried out 
for the case of global coupling, 
and clarified characters of phase transition of entrained 
oscillators\cite{Kuramoto75,Strogatz91,Daido92,Crawford95,Strogatz00}.
The case of local coupling has also been studied 
in various contexts motivated by rhythmic phenomena in biological
systems. 
They include animal gaits\cite{Collins93}, 
fish swimming\cite{Cohen82,Murray93}, 
and the peristaltic movement of 
gastrointestinal tracts\cite{Winfree80,Ermentrout84,Daido99}.
In recent years,
researches of coupled oscillators are extended to
the case of non-local (but not global) coupling\cite{Kuramoto95}
and more complicated network systems\cite{Strogatz01}.

Since such complicated networks have, in general, 
intricate boundaries and inhomogeneous couplings,
we have been found little clues to analyses of
dynamical behaviour of them.
We can, however, analyse some phase-locked solutions
for more simple and symmetric trees of couplings.
In the case of $m$-ary trees with the height $n$ and $m^n$ leaves, 
for instance, the system of coupled oscillators has some particular
solutions with identical phases in each level, and
these solutions are reduced to the phase-locked solutions for chains
of oscillators with asymmetric coupling strengths\cite{Yamada01}.
As well as complicated network structures, 
intricate dynamical processes appear in biological systems,
of which it is impossible to drag the inherent or isolated parts out.
It is one of the ways to study such systems that
we construct models from the phenomenological evidence 
and draw qualitative features of them.
When we compare the models with data observed in real systems
such as efficiency of movement and the transport volume,
we will need to know the dynamical behaviour and solutions of the
models in detail.

In this paper, 
we study chains of coupled phase oscillators.
It is the remarkable point that 
we can derive phase-locked solutions
of the systems by algebraic calculations, 
even if the coupling strengths are asymmetric or
the natural frequencies are inhomogeneous.
Moreover, 
the analytical form of the solutions is not much cumbersome, 
and thus the profile of solutions are easy to see.
From the existence conditions of the phase-locked solutions,
we can estimate transition points of 
phase-locked to phase drift states. 

We present coupled phase oscillators with asymmetric coupling
strengths in the following section,
and derive the phase-locked solutions 
in section~\ref{sec: phase lags}.
The profile of the phase-locked solutions and bifurcation phenomena
are studied for the cases of asymmetric coupling strengths 
(section~\ref{sec: asymmetry couplings}) and
of a linear frequency 
gradient (section~\ref{sec: linear frequency gradient}).
In section~\ref{sec: continuous system},
we discuss the difference analogue of the continuous system 
corresponding to the oscillator systems with asymmetric coupling 
strengths. 

\section{Coupled phase oscillators}
\label{sec: coupled phase oscillators}

A linear chain of phase oscillators with asymmetric couplings is
governed by the following equations
of phases $\theta_j$ ($j = 0, \ldots, n$):
\begin{eqnarray}
  \dot{\theta}_{0} &=& \omega_{0}
  + a_u^* h(\theta_{1} - \theta_{0}),
  \nonumber\\
  \dot{\theta}_{j} &=& \omega_{j}
  + a_d h(\theta_{j-1} - \theta_{j})
  + a_u h(\theta_{j+1} - \theta_{j}),
  \quad (j = 1, \ldots, n-1),
  \label{eq: phase0}\\
  \dot{\theta}_{n} &=& \omega_{n}
  + a_d^* h(\theta_{n-1} - \theta_{n}),
  \nonumber
\end{eqnarray}
where $\omega_j$ is the natural frequency,
$a_u$ and $a_d$ are coupling coefficients, 
and $h$ is a periodic function.
For the $0$th and $n$th oscillators, 
coupling coefficients, $a_u^*$ and $a_d^*$ are set 
according to the boundary conditions.

Introducing new variables defined by neighbouring sites,
\begin{displaymath}
  \psi_{j} := \theta_{j} - \theta_{j-1},\quad
  d_{j} := \omega_{j} - \omega_{j-1},
\end{displaymath}
we obtain $n$ equations for the phase differences:
\begin{eqnarray}
  \dot{\bm{\psi}}
  & = & \bm{d} + A_u \bm{h}_{+} - A_d \bm{h}_{-},
  \label{eq: phase}\\
  \bm{\psi} &:=& \left[ \begin{array}[c]{l}
      \psi_1 \\ \psi_2 \\ \vdots \\
      \psi_{n-1} \\ \psi_n
    \end{array} \right],
  \quad
  \bm{d} := \left[ \begin{array}[c]{l}
      d_1 \\ d_2 \\ \vdots \\
      d_{n-1} \\ d_n
    \end{array} \right],
  \quad
  \bm{h}_\pm := \left[ \begin{array}[c]{l}
      h(\pm \psi_1) \\ h(\pm \psi_2) \\ \vdots \\
      h(\pm \psi_{n-1}) \\ h(\pm \psi_n)
    \end{array} \right],
  \nonumber\\
  A_u &:=& \left[ \begin{array}[c]{rrrrr}
      - a_u^* & a_u & & & \\
      & - a_u & a_u & & \\
      & & \ddots & \ddots & \\
      & & & - a_u & a_u \\
      & & & & - a_u
    \end{array} \right],
  \nonumber\\
  A_d &:=& \left[ \begin{array}[c]{rrrrr}
      - a_d & & & & \\
      a_d & - a_d & & & \\
      & \ddots & \ddots & & \\
      & & a_d & - a_d & \\
      & & & a_d & - a_d^*
    \end{array} \right].
  \nonumber
\end{eqnarray}

In the analysis for phase locked/drift solutions shown below,
we assume $h$ is an odd function, $h(-\psi) = - h(\psi)$, 
and thereby $\bm{h}_{-} = - \bm{h}_{+}$.

\section{Phase lags}
\label{sec: phase lags}

Phase-locked solutions of eq.~\ref{eq: phase} mean all the oscillators 
have the same period 
and the phase difference is constant, $\dot{\psi}_j = 0$.
Thus they are solutions of
\begin{equation}
  A \bm{h} + \bm{d} = \bm{o},
  \label{eq: algebraic}
\end{equation}
where $A := A_u + A_d$ and $\bm{h} := \bm{h}_{+}$.
We solve formal solution of the above equations 
under the free boundary conditions (mutual entrainment), 
and one-sided coupling conditions (forced entrainment).

Fast, we study the case of free boundary conditions,
$a_u^* = a_u$, $a_d^* = a_d$, 
where the boundary oscillators and their neighbours are coupled
in the bi-directional manner.
Under these conditions, 
all of the oscillators have the same period 
\begin{equation}
  \dot\theta_j = \Omega_n / \Lambda_n,
  \label{eq: frequency}
\end{equation}
where $\Omega_n$ and $\Lambda_n$ are defined below.
From eq.~\ref{eq: algebraic}, 
we get the solution $h_j$ ($j = 1, \ldots, n$), 
the $j$th element of $\bm{h}$, 
in the form of
\begin{equation}
  a_d h_j =
  \Delta_{n-j}
  - (\Delta_n / \Lambda_n) \Lambda_{n-j},
  \label{eq: lock}
\end{equation}
or
\begin{equation}
  a_u h_j =
  \tilde{\Delta}_{j-1}
  - (\tilde{\Delta}_n / \tilde{\Lambda}_n) \tilde{\Lambda}_{j-1},
  \label{eq: lock2}
\end{equation}
where
\begin{eqnarray*}
  &&
  \Omega_j := \sum_{k=0}^{j}
  \omega_k \Bigl( \frac{a_u}{a_d} \Bigr)^{k},
  \quad
  \Lambda_j := \sum_{k=0}^{j}
  \Bigr( \frac{a_u}{a_d} \Bigr)^{k},
  \quad
  \tilde{\Lambda}_j := \sum_{k=0}^{j}
  \Bigr( \frac{a_d}{a_u} \Bigr)^{k},
  \\
  &&
  \Delta_j := \sum_{k=0}^j
  (\omega_{n-j+k} - \omega_0) \Bigr( \frac{a_u}{a_d} \Bigr)^{k},
  \quad
  \tilde{\Delta}_j := \sum_{k=0}^j
  (\omega_n - \omega_{j-k}) \Bigr( \frac{a_d}{a_u} \Bigr)^{k}.
\end{eqnarray*}
The phase difference, $\psi_j$, is obtained by inverting
eq.~\ref{eq: lock}.
When the coupling function $h$ is continuous and bounded,
solutions exist only if $\min{h} < h_j < \max{h}$ for all $j$.
The linear-stability conditions of solutions are
obtained from the eigenvalue problem of the linearized matrix,
$A \bm{h}'$, 
where the $j$th elements of $\bm{h}'$ is $h'(\psi_j)$.

We comment the cases of forced oscillations, that is,
one of the terminals is a forced oscillator.
Suppose that the $0$th oscillator is one-sided coupling to its
neighbour, $a_u^* = 0$, 
but the opposite end satisfies the free boundary condition,
$a_d^* = a_d$.
Under these conditions, 
a solution of eq.~\ref{eq: algebraic} is
$h_j = \Delta_{n-j} / a_d$.
If the boundary conditions are exchanged, that is, 
the $n$th oscillator is a forced oscillator, $a_d^* = 0$, and
the $0$th oscillator satisfies the free condition, $a_u^* = a_u$, 
a solution of eq.~\ref{eq: algebraic} is obtained in the form of
$h_j = \tilde{\Delta}_{j-1} / a_u$.

\section{Asymmetry couplings}
\label{sec: asymmetry couplings}

We study the mutual entrainment to a pacemaker with asymmetry
couplings in the present section.
We assume that all the oscillators have the same natural frequency 
but only the $0$th oscillator, the pacemaker, has the high frequency.
If we set
$\omega_0 > 0$ and $\omega_1 = \cdots = \omega_n = 0$,
eq.~\ref{eq: lock} becomes
\begin{equation}
  h_j = - \frac{\omega_0 \Lambda_{n-j}}{a_d \Lambda_n}.
  \label{eq: asym lock}
\end{equation}
The entrained frequency is obtained from eq.~\ref{eq: frequency},
\begin{equation}
  \bar{\omega}_n = \left\{ \begin{array}[c]{lr}
        \omega_0 / (1 + \cdots + \lambda^n), 
        & (a_u > a_d), \\ [0.5em]
        \omega_0 \lambda^n / (1 + \cdots + \lambda^n), & 
        (a_u < a_d),
      \end{array} \right.
  \label{eq: asym lock freq}
\end{equation}
where  $\lambda := \max(a_u, a_d) / \min(a_u, a_d)$.
We set $a_u$, $a_d$, and $h'(0)$ are positive below.
For numerical calculations shown in figures, 
we chose the coupling function as
$h(\psi) = \sin \psi + 0.1 \sin 2 \psi - 0.03 \sin 3 \psi$ 
to obtain generic results.

\begin{figure}[hbtp]
  \begin{center}
    \includegraphics[width=.3\textwidth]{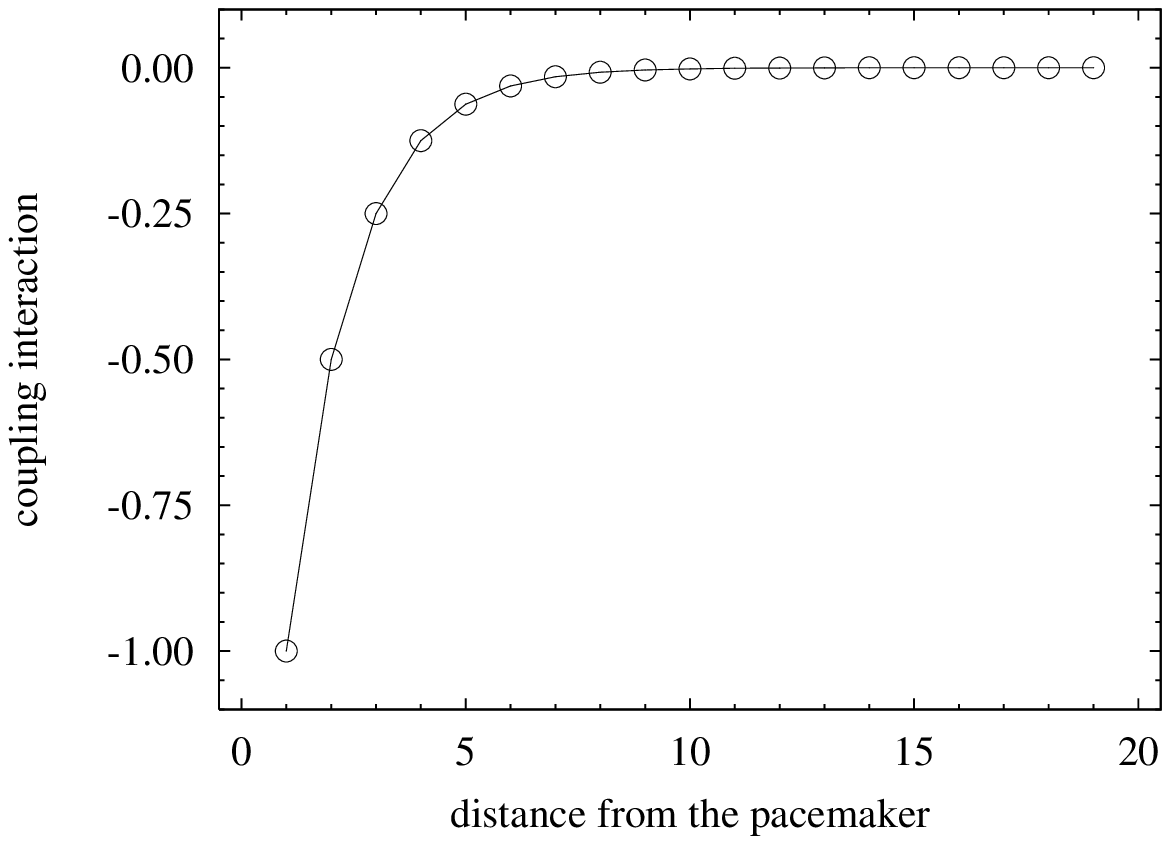}
    \includegraphics[width=.3\textwidth]{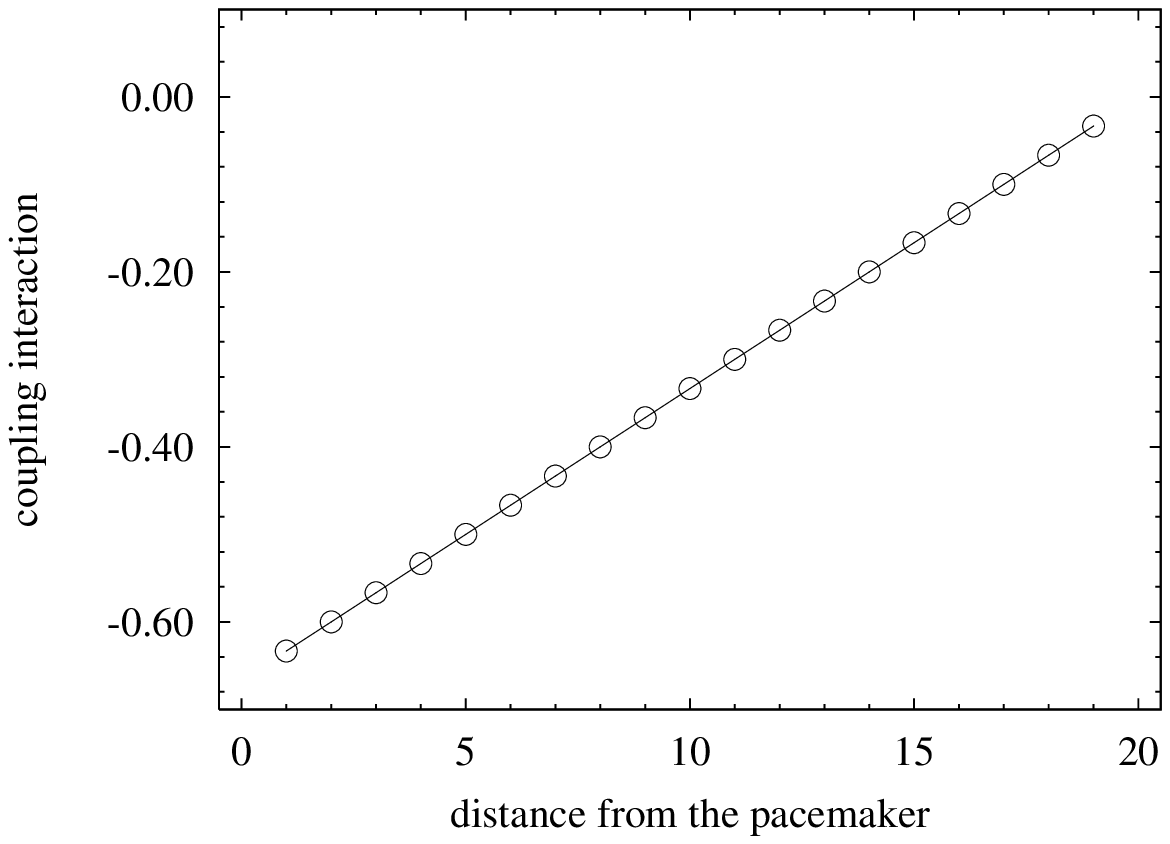}
    \includegraphics[width=.3\textwidth]{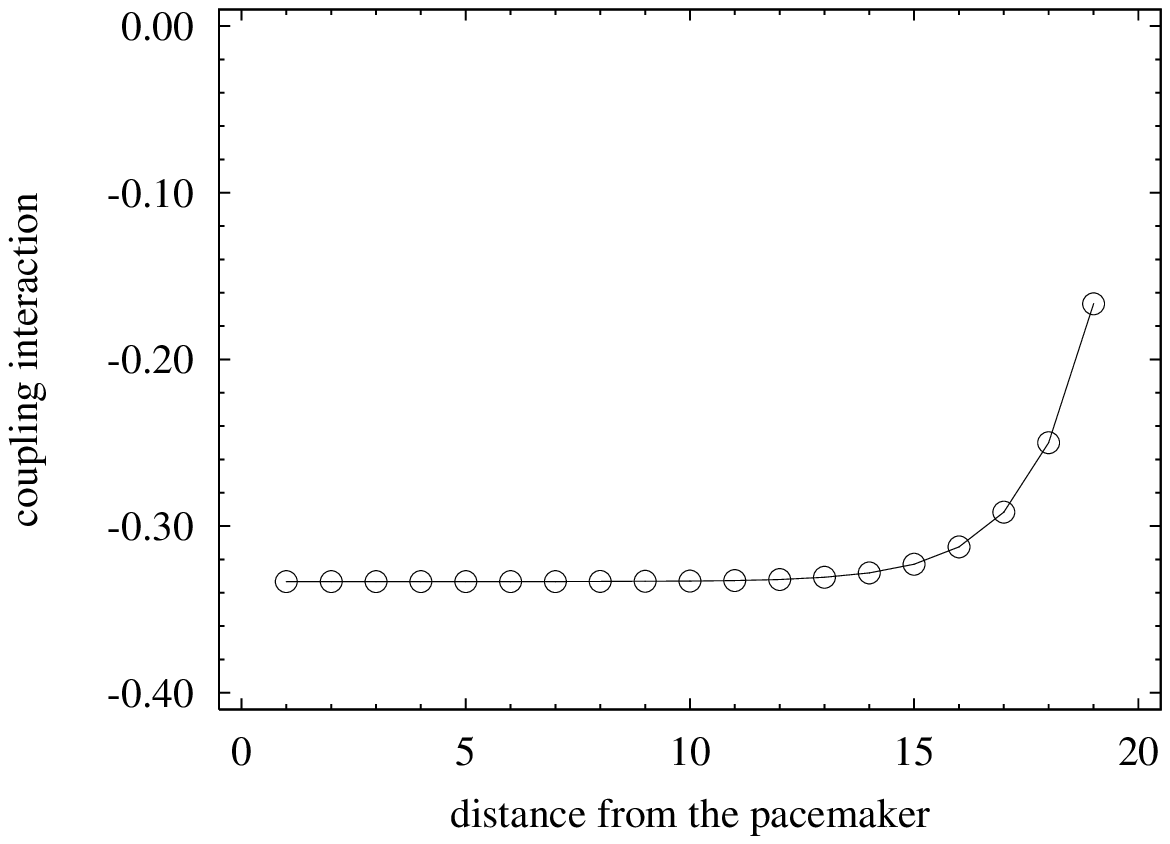}\\
    \includegraphics[width=.3\textwidth]{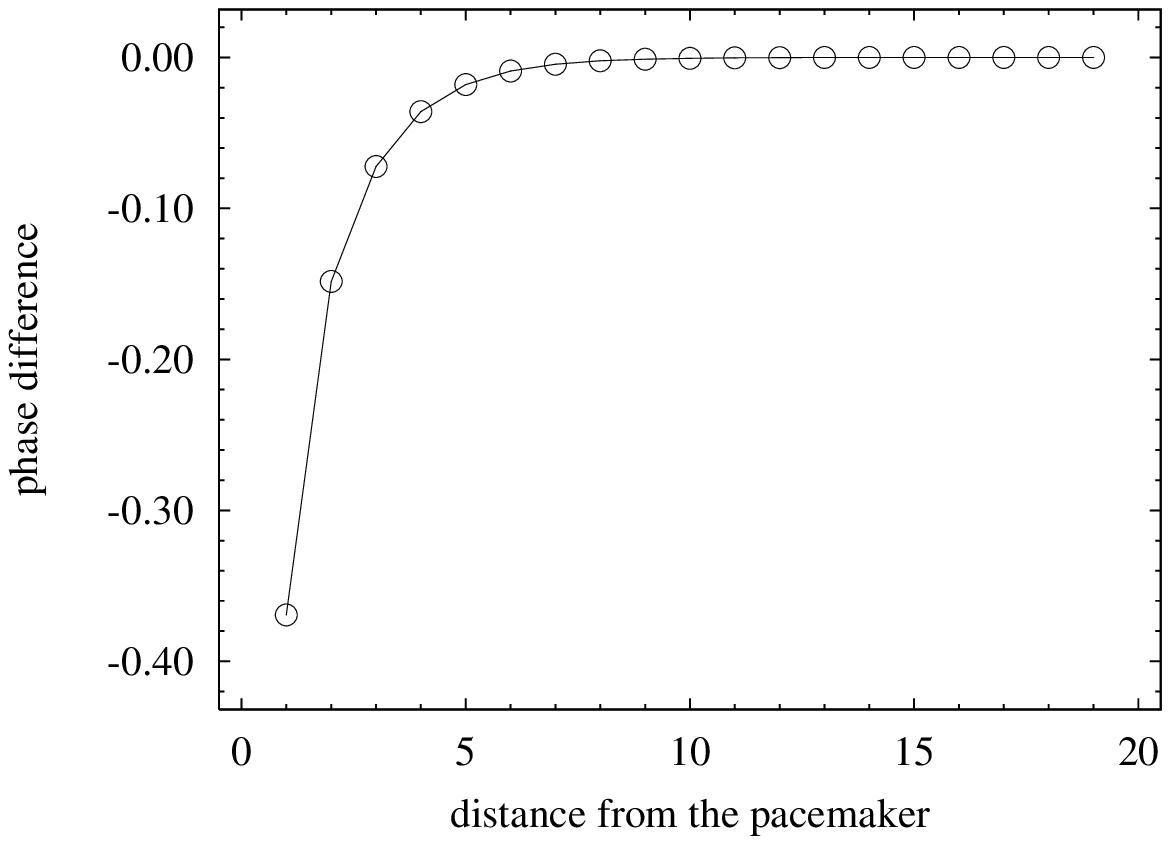}
    \includegraphics[width=.3\textwidth]{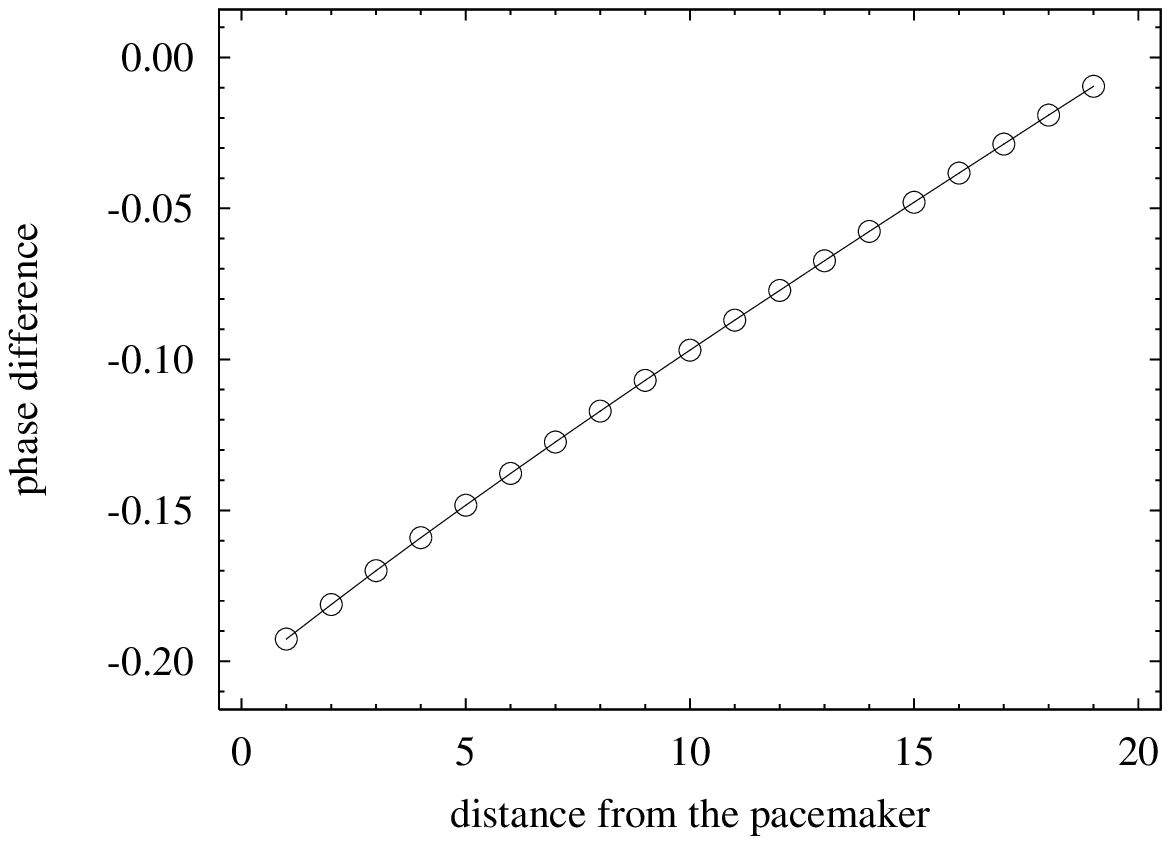}
    \includegraphics[width=.3\textwidth]{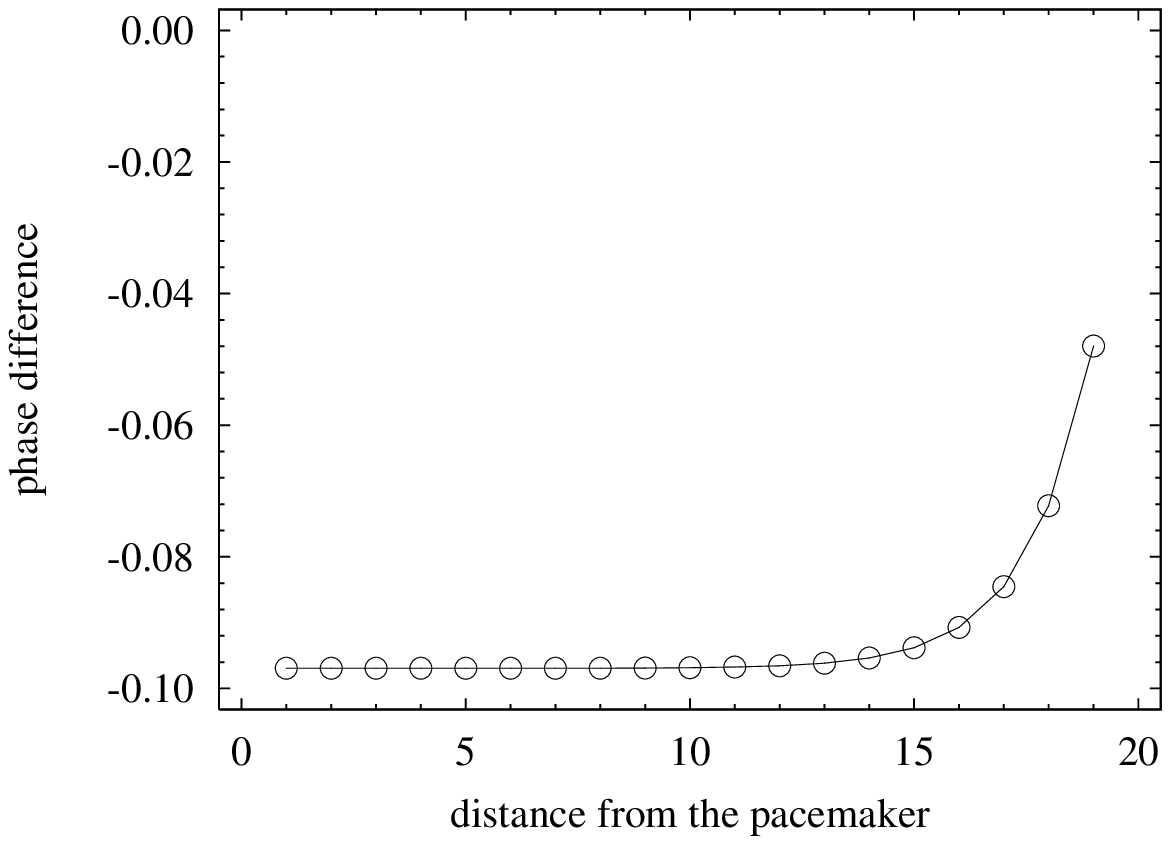}\\
    \includegraphics[width=.3\textwidth]{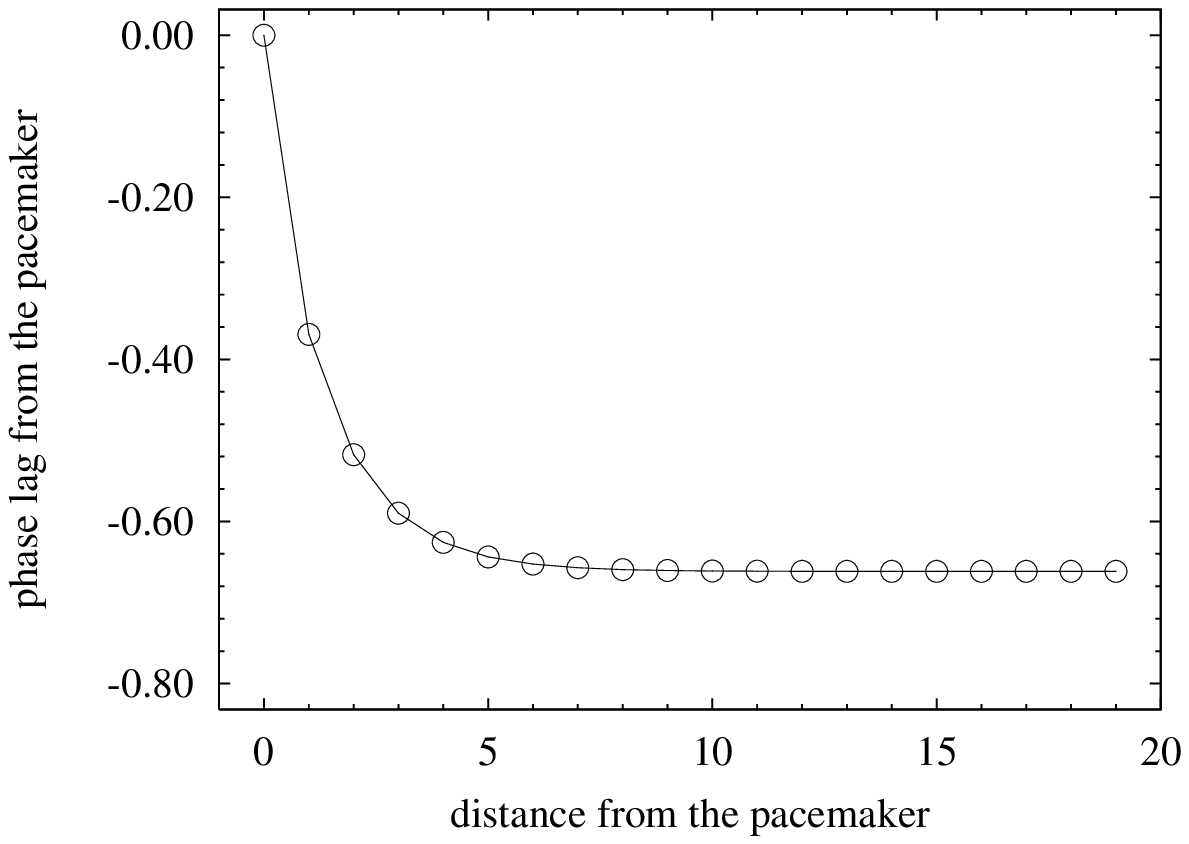}
    \includegraphics[width=.3\textwidth]{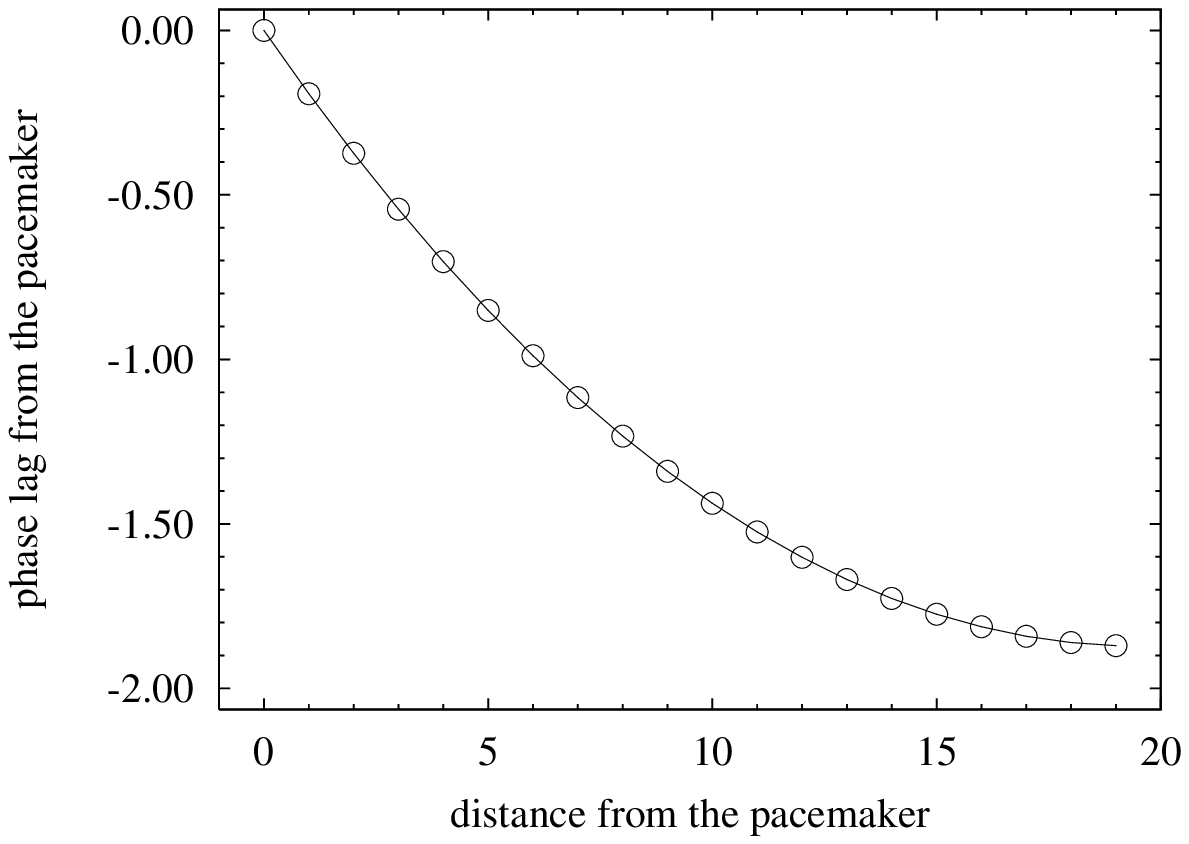}
    \includegraphics[width=.3\textwidth]{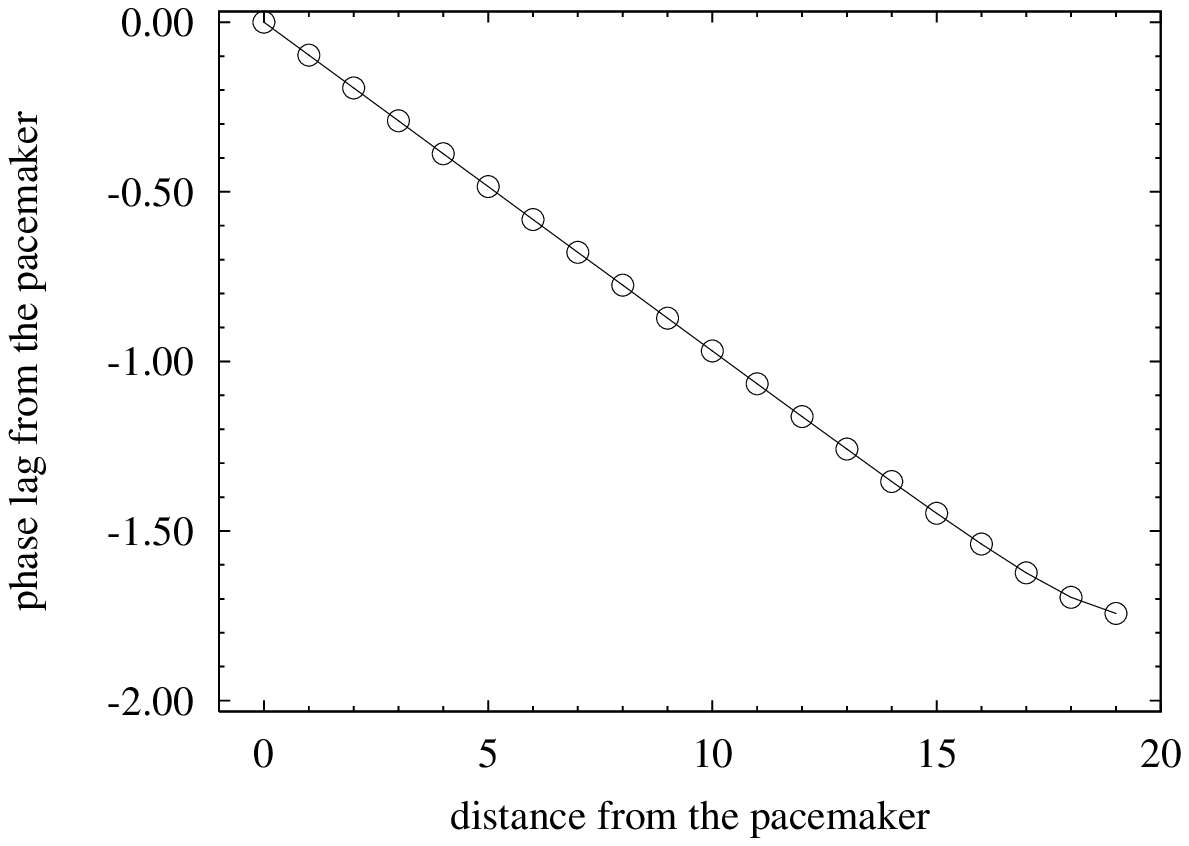}
  \end{center}
  \caption{
    The profiles of phase-locked solutions for phase oscillators 
    with asymmetric couplings: 
    the coupling interaction, $h_j$, 
    phase difference, $\psi_j$, and
    phase lag from the pacemaker, $(\theta_j - \theta_0)$, 
    where $\psi_j$ and $(\theta_j - \theta_0)$ are plotted 
    with $\pi$ units.
    The left, centre, and right columns are the cases of
    (i) $a_u / a_d = 2$ ($a_u = 1.0$, $a_d = 0.5$),
    (ii) $a_u / a_d = 1$ ($a_u = 1.5$, $a_d = 1.5$), and
    (iii) $a_u / a_d = 1/2$ ($a_u = 1.5$, $a_d = 3.0$), 
    respectively.
    Solid lines are connecting the points 
    derived eq.~\ref{eq: asym lock},
    while open circles are obtained from numerical calculations of 
    the phase equation~\ref{eq: phase0}.
    }
  \label{fig: asym lock}
\end{figure}

When the coupling strengths are asymmetric, $a_u \neq a_d$, 
the profile of $h_j$ given in eq.~\ref{eq: asym lock} 
has the exponential dependence on $j$.
In fig.~\ref{fig: asym lock},
we illustrate some features of phase-locked solutions: 
the profile of coupling interaction, phase difference and
phase lag from the pacemaker.
Calculations of these quantities are carried out
for the system of $20$ oscillators ($n = 19$) and
for three ratios of couplings, 
$a_u / a_d = 2$, $1$, and $1/2$.
If the coupling function $h(\psi)$ is approximated to the linear
function for small $\psi$,
then the dependence of the phase lag, $\theta_j - \theta_0$, 
on $j$ is almost
(i) exponential for $a_u > a_d$,
(ii) parabolic for $a_u = a_d$, and
(iii) linear for $a_u < a_d$.
Such dependence is inherited for more complicated network systems on 
trees\cite{Yamada01}.

\begin{figure}[hbtp]
  \begin{center}
    \includegraphics[width=.3\textwidth]{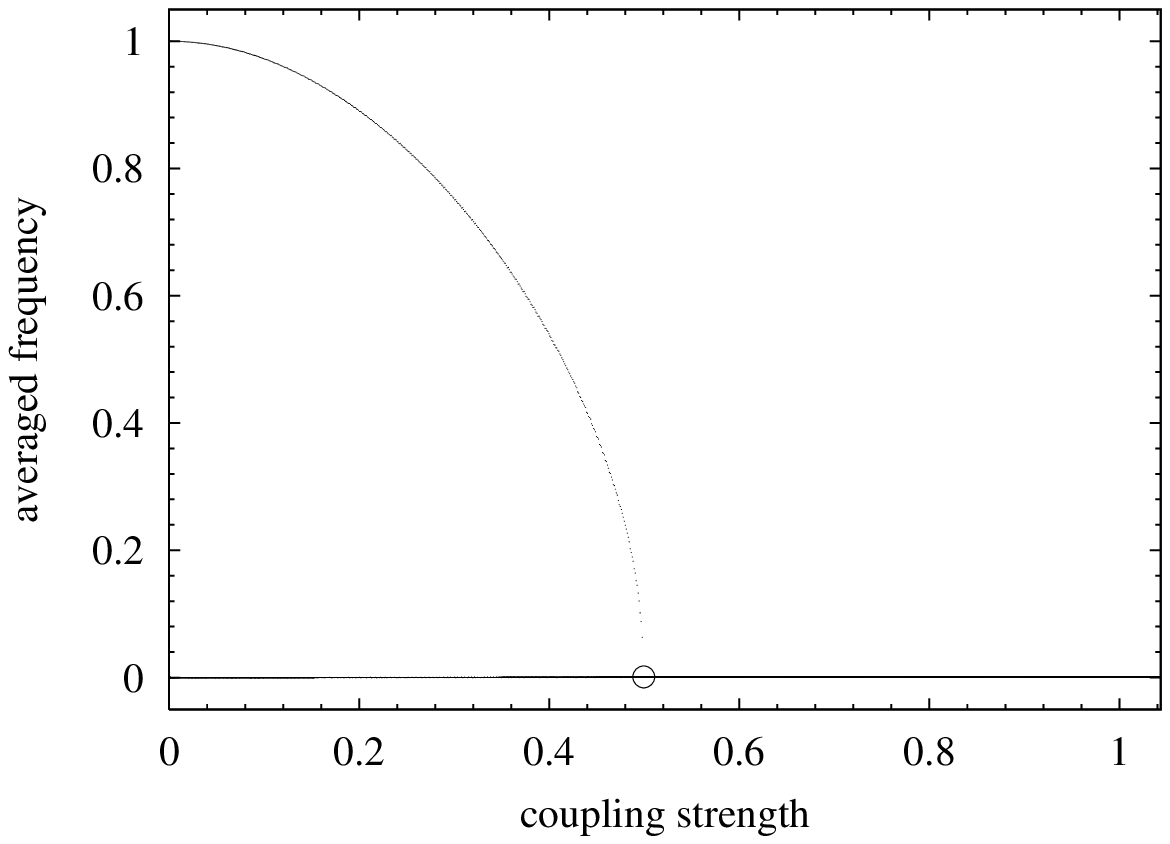}
    \includegraphics[width=.3\textwidth]{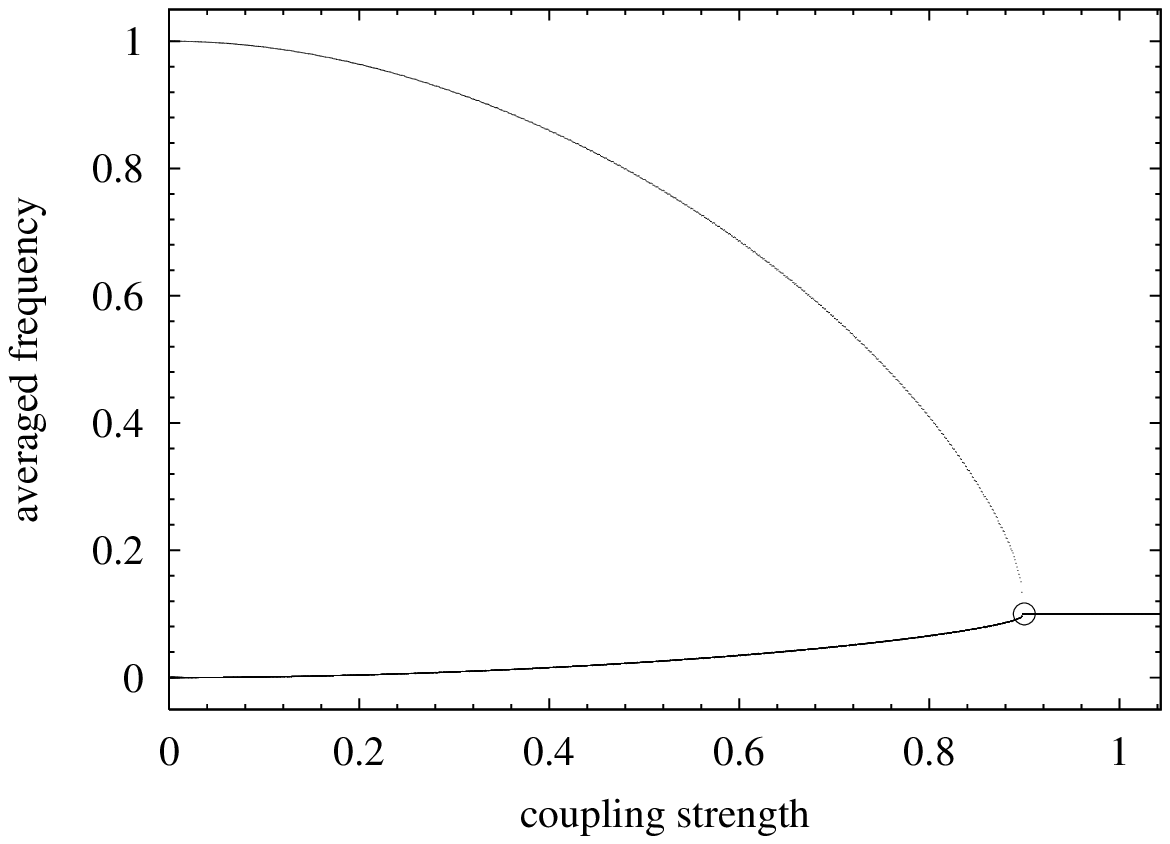}
    \includegraphics[width=.3\textwidth]{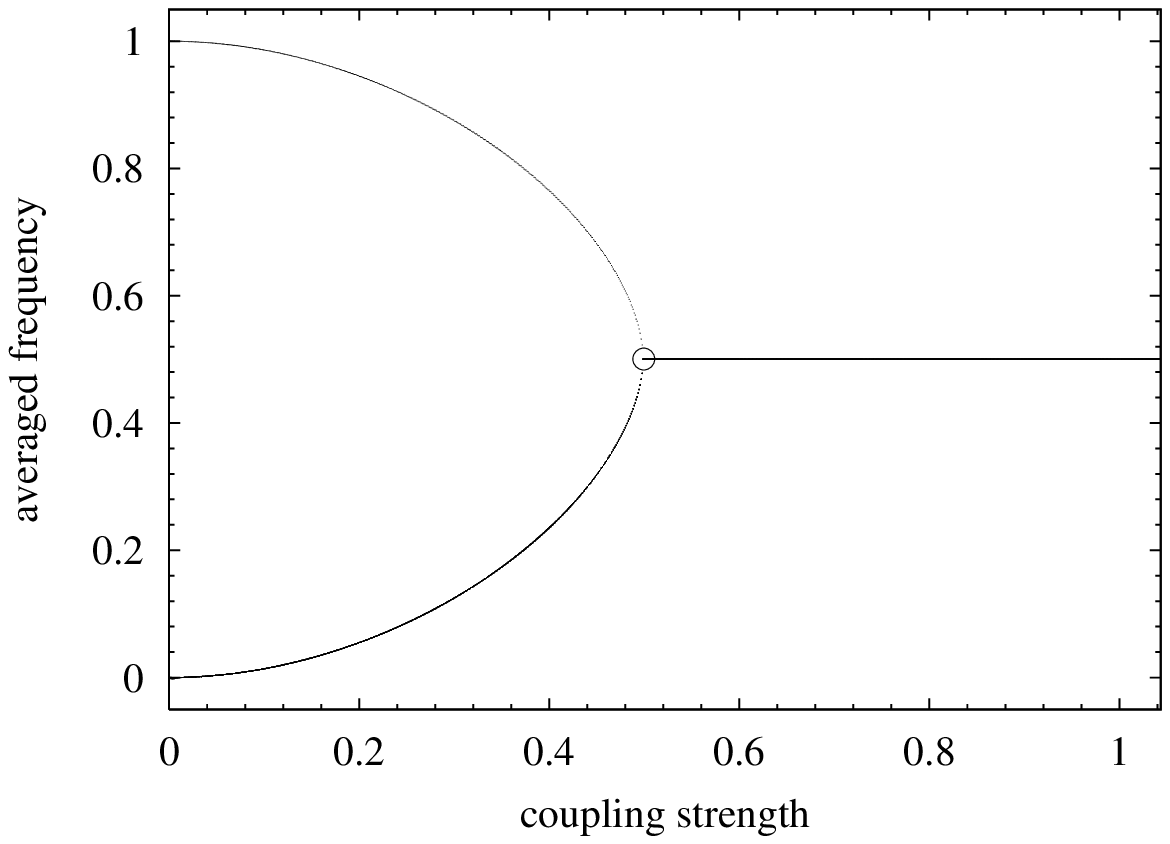}
  \end{center}
  \caption{
    The frequency diagrams for phase oscillators 
    with the asymmetric couplings.
    The left, middle, and right diagrams are the cases of
    (i) $a_u / a_d = 2$, 
    (ii) $a_u / a_d = 1$, and
    (iii) $a_u / a_d = 1/2$,
    respectively.
    The averaged frequencies $\langle \omega_j \rangle$
    are plotted against the coupling strength $\epsilon$.
    The dots are obtained from
    numerically calculations of eq.~\ref{eq: phase0}.
    The open circles denote critical points
    of the existence condition for phase-locked solutions.
    }
  \label{fig: asym bif}
\end{figure}

For small coupling strengths,
the phase-locked state is broken and 
the phase of each oscillator drifts from the phase of the pacemaker.
Such a phase drift solution has two regions of the
frequency-entrainment since the pacemaker and other oscillators have
different natural frequencies.
The phase drift occurs between these two regions.
Fig.~\ref{fig: asym bif} shows the transition between phase-locked and
phase drift solutions for the three ratios of $a_u$ and $a_d$. 
Each system has $10$ oscillators ($n = 9$).
The averaged frequency of the $j$th oscillator and the coupling
strength are defined by
\begin{equation}
  \langle \omega_j \rangle := \lim_{t \to \infty} 
  \bigl( \theta_j(t) - \theta_j(0) \bigr) \big/ t,
  \quad
  \epsilon := \min (a_u, a_d) \max h.
  \label{eq: averaged frequency}
\end{equation}
In fig.~\ref{fig: asym bif},
$\langle \omega_j \rangle$ is obtained from numerical integrations of
eq.~\ref{eq: phase0} with the step width in $\epsilon$, 
$\Delta \epsilon = 10^{-3} \times \max h$.
We can estimate the transition point of phase-locked and
phase drift solutions from the necessary condition for the
existence of the phase-locked solutions,
\begin{displaymath}
  \epsilon > \epsilon_n :=
  \omega_0 \frac{1 + \cdots + \lambda^{n-1}}{1 + \cdots + \lambda^n}.
\end{displaymath}
The critical point $(\epsilon_n, \bar{\omega}_n)$ is plotted by
the open circle in fig.~\ref{fig: asym bif}.

Numerical calculations show that 
phase drift states always consist of two regions: 
the pacemaker and the other oscillators.
It implies that 
if the pacemaker can entrain its neighbouring oscillator,
all oscillators will be entrained to the pacemaker, 
but if not so, 
no oscillators entrained to the pacemaker.
It is also possible to estimate this entrainment feature from
the existence condition of the phase-locked solutions.
We assume that the pacemaker is entraining $j$ oscillators.
Then we divide the system into two sets of coupling oscillators:
(i) upper $(j+1)$ oscillators including the pacemaker, and
(ii) lower $(n-j)$ oscillators.
For the upper set (i),
the phase-locked solution exist only if
\begin{equation}
  \omega_0 \frac{1 + \cdots + \lambda^{j-1}}{1 + \cdots + \lambda^{j}}
  < \epsilon.
  \label{eq: necessary}
\end{equation}
For the lower set (ii), 
we add one oscillator with the entrainment frequency of the set (i)
to the upper side as the pacemaker.
If the coupling strength $\epsilon$ satisfies
\begin{equation}
  \epsilon < \bar{\omega}_j
  \frac{1 + \cdots + \lambda^{n-j-1}}{1 + \cdots + \lambda^{n-j}},
  \label{eq: safficient}
\end{equation}
then the phase drift solution exists 
for the set (ii) with the pacemaker. 
Since no value of the coupling strength $\epsilon$ satisfies
both of conditions~\ref{eq: necessary} and~\ref{eq: safficient},
oscillators entraining to the pacemaker 
are all or none.

\section{Linear frequency gradient}
\label{sec: linear frequency gradient}

Daido~\cite{Daido99} has studied chains of coupled phase oscillators
with a linear gradient of natural frequencies by numerical
calculations. 
He has shown some scaling properties in the frequency diagram, 
a plot of averaged frequencies against the coupling strength.
In the present section,
we attempt to derive some of them from eq.~\ref{eq: lock}.

We assume that 
the natural frequency of the $j$th oscillator is $\omega_j := j / n$ 
and coupling strengths are symmetric $a := a_u = a_d$.
Then a phase-locked solution is obtained from eq.~\ref{eq: lock} 
in the form of
\begin{equation}
  h_j = \frac{\delta}{a} \frac{j (n + 1 - j)}{2},
  \label{eq: linear lock}
\end{equation}
where $\delta := 1 / n$ is the difference of natural frequency between 
neibouring oscillators.
The entrained frequency of this solution is
\begin{equation}
  \bar{\omega}_n = \delta \frac{n}{2} \Bigl( = \frac{1}{2} \Bigr).
  \label{eq: linear lock freq}
\end{equation}
We set $a$ and $h'(0)$ are positive, and
chose the coupling function as the previous section
for numerical calculations shown in figures below.

\begin{figure}[hbtp]
  \begin{center}
    \includegraphics[width=.3\textwidth]{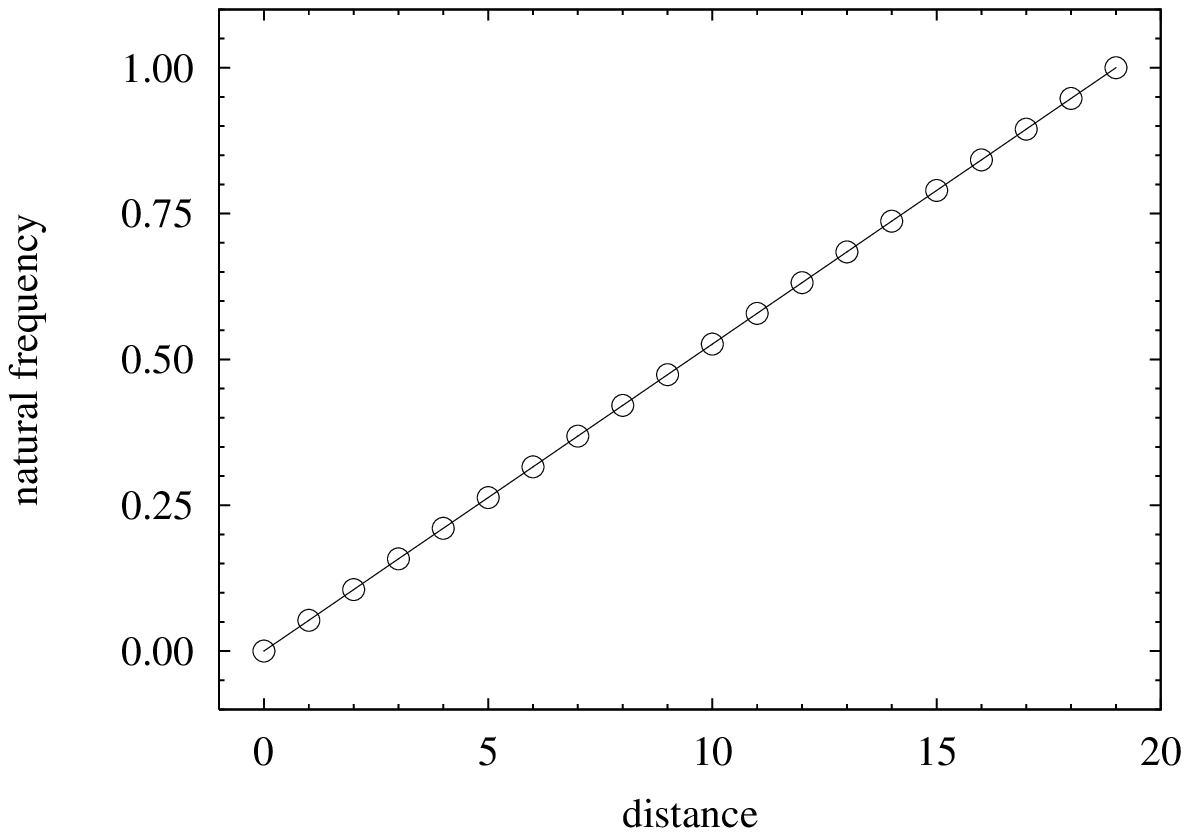}
    \includegraphics[width=.3\textwidth]{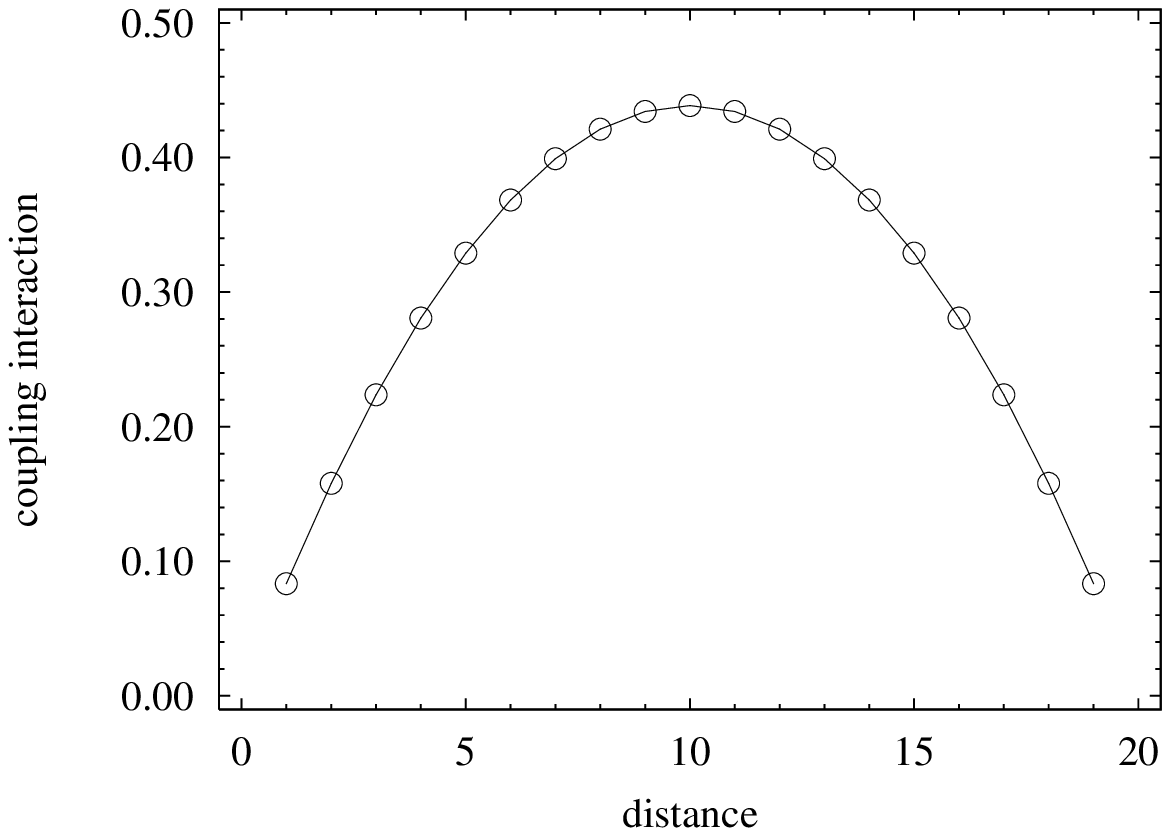}\\
    \includegraphics[width=.3\textwidth]{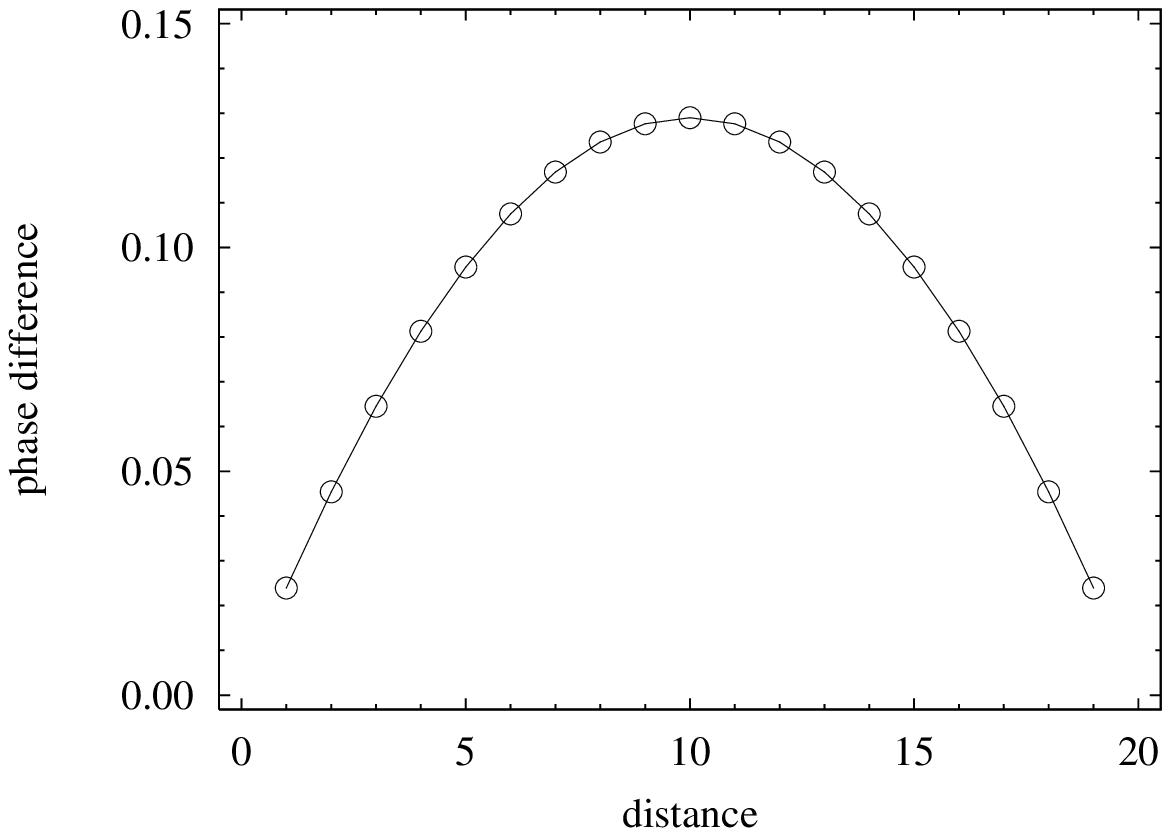}
    \includegraphics[width=.3\textwidth]{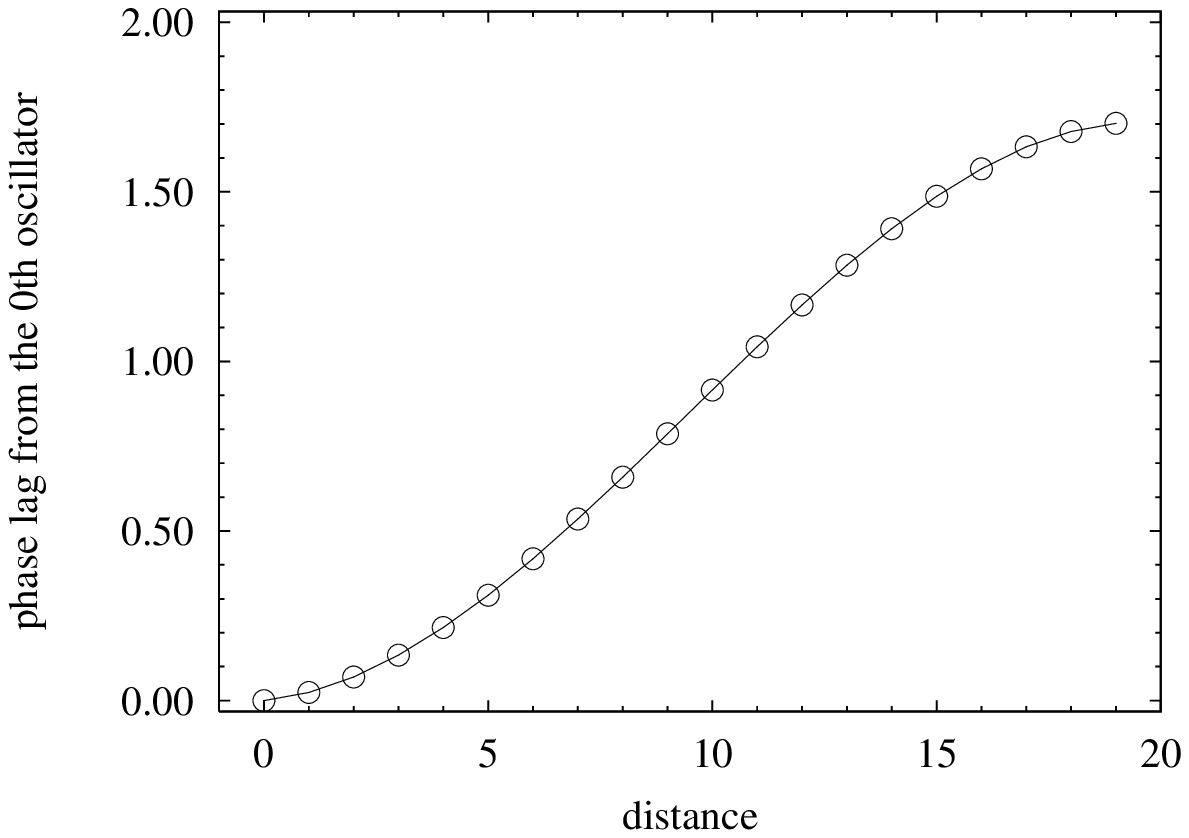}
  \end{center}
  \caption{
    The profiles of phase-locked solutions for phase oscillators 
    with linear frequency gradient:
    the coupling interaction, $h_j$, 
    phase difference, $\psi_j$, and
    phase lag from the 0th oscillator, $(\theta_j - \theta_0)$.
    The upper left shows the natural frequency of each oscillator.
    }
  \label{fig: linear lock}
\end{figure}

We note that $h_j$ takes the maximal value at the centre
of the chain of entrained oscillators from eq.~\ref{eq: linear lock}.
If the phase difference $\psi_j$ is small enough to approximate that 
the coupling function $h(\psi)$ has the linear dependence on $\psi$, 
$\psi_j$ has the convex dependence on $j$ as well as $h_j$.
Then the phase lag $(\theta_j - \theta_0)$ has an inflection point at 
the centre of the chain.
Figure~\ref{fig: linear lock} shows the profile of phase lags and other
variables. 

When the coupling strength is diminished,
the entrained region is divided into some clusters of
frequency-entrained oscillators, 
called frequency plateaus\cite{Winfree80,Ermentrout84,Daido99}.
In the limiting case that the coupling strength is vanishing, 
entrained domains of oscillators will split into each oscillator.

Although the frequency diagram shows fine and complicated bifurcation
structure\cite{Daido99}, 
we can estimate the approximate arrangement of transition points 
by means of the necessary conditions for existence of phase-locked
solutions. 
To illustrate the frequency diagram, 
we define the averaged frequency of the $j$th oscillator and
the coupling strength as eq.~\ref{eq: averaged frequency}.
We pick out the frequency plateau with $m$ oscillators from 
the $(n + 1)$-oscillator system, 
and consider these oscillators as the isolated system with the free
boundary conditions in both sides.
Phase-locked solutions of the $m$-oscillator system exist only if
\begin{equation}
  \epsilon > \epsilon_m := \left\{ \begin{array}[c]{lr}
        (\delta / 8) m^2, & (m : \mbox{even}),\\[0.5em]
        (\delta / 8) (m^2 - 1), & (m : \mbox{odd}).
      \end{array} \right.
  \label{eq: m-necessary}
\end{equation}
The entrained frequency is the average of natural frequencies,
\begin{eqnarray}
  \bar{\omega}_{k, k + m - 1}
  &=& (\omega_k + \cdots + \omega_{k + m - 1}) / m 
  \nonumber\\
  &=& \delta (m + 2 k - 1) / 2,\quad
  (k = 0, \ldots, n-m).
  \label{eq: m-omega}
\end{eqnarray}

\begin{figure}[hbtp]
  \begin{center}
    \includegraphics[width=.8\textwidth]{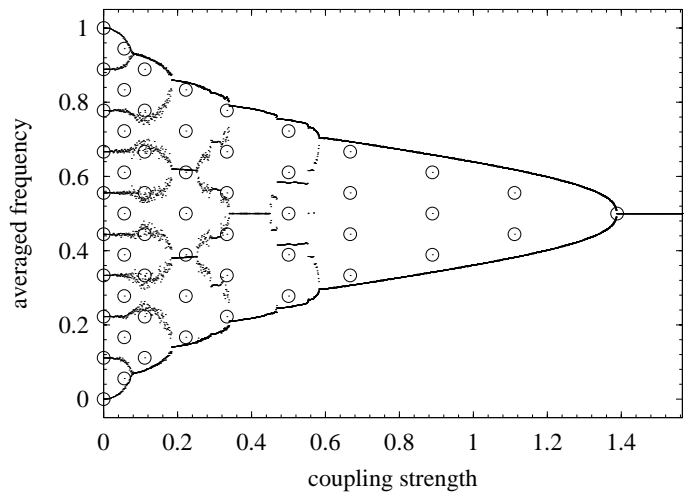}
  \end{center}
  \caption{
    The frequency diagram for phase oscillators with
    the linear gradient of natural frequencies.
    The diagram is plotted 
    the averaged frequencies $\langle \omega_j \rangle$
    against the coupling strength $\epsilon$.
    Averaged frequencies plotted by dots are numerically calculated.
    The open circles denote critical points for the existence
    condition of phase-locked solutions 
    with $m$ oscillators ($m = 1, \ldots, n+1$).
    }
  \label{fig: linear bif}
\end{figure}

In fig.~\ref{fig: linear bif}, 
we put the approximate critical point 
$(\bar{\omega}_{k, k+m-1}, \epsilon_m)$ on the frequency diagrams for
the system of 10 oscillators ($n = 9$).
Some of approximate points are located near the transition points, 
but others have no transition points around them.
The latter points indicate that it is impossible for the system 
to have some states such as
small plateaus located near the boundaries or
extremely asymmetric arrangements of plateaus.
We obtain the scaling behaviour of 
$\bar{\omega}_{0, m-1}$ to $\epsilon_m / \epsilon_{n+1}$ from 
eqs.~\ref{eq: m-necessary} and~\ref{eq: m-omega} as
\begin{equation}
  \bar{\omega}_{0, m-1}
  \sim \frac{1}{2} \left(\frac{\epsilon_m}{\epsilon_{n+1}}\right)^{1/2}
  \label{eq: scaling}
\end{equation}
in the limit of $n \to \infty$.

\section{Discussion}
\label{sec: continuous system}

We can consider the phase-difference system~\ref{eq: phase} is 
derived from some continuous partial differential equation by means of 
the difference analogue.
If the coupling function $h$ is assumed to be the odd function,
the continuous system corresponding to eq.~\ref{eq: phase} becomes
\begin{equation}
  \partial_t \psi + \Gamma \partial_x h(\psi)
  = d(x) + D \partial_x^2 h(\psi),
  \label{eq: continuous}
\end{equation}
where $x$ is the spatial coordinate. 
Coefficients $\Gamma$ and $D$ are defined by
\begin{displaymath}
  \Gamma := (a_d - a_u) \Delta x,\quad
  D := \frac{a_d + a_u}{2} (\Delta x)^2
\end{displaymath}
for the difference interval $\Delta x$. 
Here we took the central difference for the differential of first
order. 
In the continuous system~\ref{eq: continuous},
entrained solutions are obtained from the ordinary differential
equation in the form of
\begin{equation}
  D \frac{\d^2 h}{\d x^2} - \Gamma \frac{\d h}{\d x}
  + d(x) = 0.
  \label{eq: ode}
\end{equation}

When all of natural frequencies are identical, $d(x) \equiv 0$,
solutions of eq.~\ref{eq: ode} are
\begin{displaymath}
  h(x) = \left\{ \begin{array}[c]{lr}
      c + b \exp (\Gamma x / D), & (\Gamma \neq 0), \\[0.5em]
      c + b x, & (\Gamma = 0),
    \end{array} \right.
\end{displaymath}
where $b$ and $c$ are constants defined by boundary conditions.
These solutions are consistent with 
eq.~\ref{eq: asym lock} and fig.~\ref{fig: asym lock}.
In another case corresponding to the linear frequency gradient,
$d(x) = \delta$ (const.), a solution of eq.~\ref{eq: ode} are obtained
in the form of
\begin{displaymath}
  h(x) = c + b x - (\delta / 2 D) x^2.
\end{displaymath}
This is also consistent with 
eq.~\ref{eq: linear lock} and fig.~\ref{fig: linear lock},
if we set arbitrary constants $b$ and $c$ in proper values.

Here we give more examples.
When the both cases mentioned above are combined, 
that is,
the natural frequency has the linear gradient in space, 
$d(x) = \delta$, 
and coupling strengths are asymmetric, $\Gamma \neq 0$, 
a solution of eq.~\ref{eq: ode} is
\begin{displaymath}
  h(x) = c + (\Gamma \delta / D) x + b \exp (\Gamma x / D).
\end{displaymath}
Even if $\Gamma = 0$,
we can obtain the similar dependence of $h$ on $x$ from the system
in which the natural frequency has the exponential gradient in space
as $d(x) = \gamma \exp \bigl(\gamma (x-L) \bigr)$ 
($L$ is the system size). 
A solution of this system is
\begin{displaymath}
  h(x) = c + b x - (1 / D \gamma) \exp \bigl( \gamma (x - L) \bigr).
\end{displaymath}

In the last, 
we comment on the case that
the coupling functions is divided into the odd and even parts.
In general, 
the division of a function $h(x)$ is carried out in the form of
\begin{eqnarray*}
  h(x) &=& h_{\rm odd}(x) + h_{\rm even}(x),
  \\
  &&
  h_{\rm  odd}(x) := \frac{h(x) - h(-x)}{2},
  \quad
  h_{\rm even}(x) := \frac{h(x) + h(-x)}{2}.
\end{eqnarray*}
The system of coupled phase oscillators~\ref{eq: phase} is derived 
from the following continuous system
by the discretization of space
\begin{eqnarray}
  &&\partial_t \psi
  + \Gamma_{\rm  odd} \partial_x h_{\rm  odd}(\psi)
  - \Gamma_{\rm even} \partial_x h_{\rm even}(\psi)
  \nonumber\\
  &&= d(x)
  + D_{\rm  odd} \partial_x^2 h_{\rm  odd}(\psi)
  - D_{\rm even} \partial_x^2 h_{\rm even}(\psi),
  \label{eq: odd-even}
\end{eqnarray}
where coefficients are defined as
\begin{eqnarray*}
  &&
  \Gamma_{\rm  odd} := (a_d - a_u) \Delta x,\quad
  \Gamma_{\rm even} := (a_d + a_u) \Delta x
  \\&&
  D_{\rm  odd} := \frac{a_d + a_u}{2} (\Delta x)^2,\quad
  D_{\rm even} := \frac{a_d - a_u}{2} (\Delta x)^2.
\end{eqnarray*}
We note that the coefficients of advection and diffusion terms are
almost exchanged between the odd and even parts,
\begin{displaymath}
  D_{\rm odd} = (\Delta x / 2) \Gamma_{\rm even}, \quad
  D_{\rm even} = (\Delta x / 2) \Gamma_{\rm odd}.
\end{displaymath}
It is also possible to obtain the phase-locked solutions 
from eq.~\ref{eq: odd-even} as above,
but we do not mention them here.

\section{Conclusions}

We obtained the phase-locked solutions of coupled phase oscillators 
with asymmetric coupling strengths or inhomogeneous natural
frequencies. 
These solutions show the different profiles of phase lags between
the three types of ratios of the upward to downward
coupling strength.
The phase differences obtained in analytical form,
must be applicable to consider the efficiencies of 
biological motions and transportations 
based on the collective oscillation.
Moreover, inherent structures of the real systems,
the asymmetry of couplings, for example,
are suggested by the spatio-temporal patterns
observed in the systems.

From the existence conditions of phase-locked solutions,
we derived transition points from phase-locked to 
phase drift states.
We also obtained that
the length of entrained region to the pacemaker is approximately
zero or the system size.
However,
for the discrete time-dependent Ginzburg-Landau equations,
the number of entraining oscillators to the pacemaker depends on the 
coupling strength, and the sequence of bifurcations is observed
numerically\cite{Yamada01}.

Next, we showed the profiles of phase-locked solutions for
the linear frequency gradient.
The phase differences take the maximal absolute value around the
centre of the chain system.
This means that the phase drift tends to occur around the centre 
when the coupling strengths are diminished.
As well as the case of the asymmetric coupling strengths,
we estimate the approximate transition points in the frequency diagram
from the existence conditions of phase-locked solutions,
and derive some scaling properties.


\section*{Figures}

\end{document}